\documentclass[10pt]{article}
\begin{document}
\title{Generalisations of the recent Pusey-Barrett-Rudolph theorem for statistical models of quantum phenomena}
\author{Michael J. W. Hall\\
Centre for Quantum Computation and Communication\\ Technology (Australian Research Council);\\
Centre for Quantum Dynamics, Griffith University,\\ Brisbane, QLD 4111, Australia}

\maketitle



\begin{abstract}
Pusey, Barrett and Rudolph (PBR) have recently given a completely novel argument that restricts the class of possible models for quantum phenomena \cite{pbr}.  In these notes the assumptions used by PBR are considerably weakened, to  further restrict the class of possible models.  The `factorisability' assumption used by PBR  is replaced by a far weaker `compatibility' assumption for uncorrelated quantum subsystems which, moreover, does not require the assignation of separate underlying properties to each subsystem (i.e, reductionism). Further, it is shown that an assumption of measurement independence may be dropped, to obtain a related result having the same experimental significance (at the expense of a weaker conceptual significance). This is a remarkable feature of the  PBR approach, given that Bell inequalities, steering inequalities and Kochen-Specker theorems all require an assumption of this type.
\end{abstract}


\section{Introduction}

Various `no-go' theorems exist for models of quantum phenomena, based on various more or less plausible assumptions for the structure of such models.  Such theorems include derivations of Bell inequalities, Kochen-Specker theorems, and steering inequalities \cite{bell,ks,steer}, and support a longstanding view that quantum mechanics is more or less implausible - even `shocking' according to Bohr \cite{bohr}. This has led not only to much philosophical discussion on which assumption(s) should be relaxed, but also to surprising applications of what might be termed `quantum implausibility', such as quantum cryptography, quantum steering, and quantum computation.

Very recently, Pusey, Barrett and Rudolph (PBR) have given a completely novel `no-go' theorem \cite{pbr}, which demonstrates that, under certain assumptions, distinct pure quantum states must have disjoint sets of underlying properties.   In particular, partially overlapping wave functions {\it cannot} be considered as partially overlapping ensembles of underlying properties, in any model that satisfies the PBR assumptions.  Rather, the corresponding ensembles must be nonoverlapping, implying that the wave functions must be considered precisely just as `real' or `physical' as the underlying properties themselves.

The strength of the PBR theorem is determined by the strength of the assumptions it relies on.  Hence, it is of obvious interest to try to weaken these assumptions as much as possible.  This is the aim of the present notes.

For example, the`factorisability' assumption used by PBR  can be replaced by a far weaker `compatibility' assumption for the preparations of uncorrelated quantum states.  Moreover, this weaker assumption does not require the underlying model to be `reductionist': properties describing a quantum system are not assumed to partition into properties describing its subsystems, even when these subsystems appear to be uncorrelated at the quantum level.  Further, for the case of reductionist models, one may instead replace `factorisability' by a weaker `local compatibility' assumption.

It is also shown that an assumption of measurement independence implicitly used by PBR (often justified in the literature by an appeal to `free will' of experimenters to choose measurement settings), may be dropped, to obtain a related result having the same experimental significance - albeit it at the expense of a weaker conceptual significance. This is a remarkable and apparently unique feature of the  PBR approach, arising from its consideration of a {\it single} measurement procedure for any given pair of pure states.  In contrast, Bell inequalities, steering inequalities and Kochen-Specker theorems all require two or more incompatible (i.e., counterfactual) measurement settings, so that an assumption equivalent to or stronger than measurement independence must be made \cite{relaxed}.

In the following section the relevant notation is introduced for describing general underlying models of quantum phenomena, largely following that of PBR in Ref.~\cite{pbr}, and a brief overview is given of the assumptions required to obtain generalisations of the PBR theorem.  Section 3 defines and discusses these assumptions in detail, and two corresponding theorems that strengthen the original PBR theorem, corresponding to whether or not an assumption of measurement independence is made, are obtained and discussed in Section 4.

\section{Underlying models}

Let $\lambda$ describe any underlying properties following some preparation procedure $P_\Psi$ of a pure quantum state $\Psi$, and before some measurement procedure $M$.  Hence, Bayes theorem tells us that the probability of measurement outcome $M=m$ is
\begin{equation} \label{bayes}
     p(m | M, P_\Psi) = \int d\lambda\, p(m | M, P_\Psi,\lambda) \, p(\lambda | M, P_\Psi)
\end{equation}
It is not necessary to argue here whether the properties described by $\lambda$ are `real' or `not real' - what matters is that Bayes theorem is satisfied.  This is the case, for example, whether or not one has an objective or subjective view of probability. However, it can be convenient to consider $\lambda$ as describing the fundamental beables of the model.

Note that it is possible for the wave function $\Psi$ to be one of the properties (or possibly the only property) described by $\lambda$. The aim of the PBR approach is to give reasonable assumptions under which this must be the case.  In particular, under these assumptions, $\lambda$ must provide sufficient information to uniquely reconstruct $\Psi$ \cite{pbr}(up to a global phase factor).

It is important to note that $P_\Psi$ may refer to preparation of a composite or a single quantum system, and that $M$ may refer to a one-system or a multi-system measurement procedure. One only requires that one can carry out corresponding well-defined preparation and measurement procedures, e.g., by pressing buttons marked `$P_\Psi$' and `$M$' on respective apparatuses in the laboratory.  It is possible that these preparation and measurement procedures are not suitably `matched' - e.g., if $M$ requires a tensor product input state but $\Psi$ is a single system state.  In such a case the measurement will give a `null' result.  However, the only implication for Eq.~(\ref{bayes}) is that the left hand side of Eq.~(\ref{bayes}) may not correspond to a quantum probability of the simple form $\langle\Psi|M_m|\Psi\rangle$.  The parameter $\lambda$ on the right hand side nevertheless describes the corresponding underlying properties conditional on $P_\Psi$ and $M$.

Two assumptions will be required to obtain Theorem 1 in section 4 - one for each of the factors in the integral in Eq.~(\ref{bayes}).  These may be broadly stated as
\begin{list}{}{}
\item (i) {\it Statistical Completeness}: the properties described by $\lambda$ contain sufficient information to generate the outcome statistics. 
\item (ii) {\it (Local) Compatibility}: if $\lambda$ is compatible with several pure quantum states, then it is compatible with the tensor product of these states. 
\end{list}

To obtain Theorem 2 in section 4, a further assumption is required:  
\begin{list}{}{}
\item (iii) {\it Measurement Independence}: the properties described by $\lambda$ are uncorrelated with the choice of measurement $M$.
\end{list}

Note that, to derive their theorem, PBR use assumption (i) as above, but make a stronger `factorisability' assumption in place of assumption (ii). Further, while not explicitly discussed by PBR, their theorem also relies on assumption (iii), as discussed further below.  

The above assumptions are carefully defined and discussed in section 3, followed by the derivations and discussion of the corresponding Theorems 1 and 2 in section 4.

\section{The assumptions}

\subsection{Statistical completeness}

As per assumption (i) above, statistical completeness requires that the properties $\lambda$ contain sufficient information to generate the outcome statistics. Thus, the only effect of the preparation $P_\Psi$ on the measurement statistics is via the properties described by $\lambda$, i.e., \begin{equation} \label{comp}
    p(m | M, P_\Psi, \lambda) = p(m | M, \lambda).         
\end{equation}

The labelling of Eq.~(\ref{comp}) as `statistical completeness' is justified in that once $\lambda$ has been specified, any knowledge about $P_\psi$ is redundant for calculating the outcome probability distribution - all relevant information is carried by $\lambda$. Note that this in fact always holds, for example, in the following (overlapping) cases:

\begin{list}{}{}
\item
(a)  Deterministic models, in which knowledge of $\lambda$ completely determines the outcome of any measurement, i.e., $p(m|M,\lambda)\in \{0,1\}$.
\item
(b) Models which are maximally complete, i.e., $\lambda$ represents the maximum possible knowledge about the system prior to measurement. Thus, any extra information, including details of the preparation procedure, is redundant, and equation (2) must hold. Such models include deterministic models as a special case.
 \item
(c) The standard Hilbert space model, in which  $\lambda$ is the wave function $\Psi$ itself, with $p(m|M,P_\Psi,\lambda)$ in Eq.~(\ref{bayes}) given by $\langle\Psi|M_m|\Psi\rangle$ (for `matched' $P_\Psi$ and $M$ as per Section 2), and $p(\lambda| M, P_\Psi)$ by  $\delta(\lambda - \Psi)$.  Note that in the Copenhagen and many-worlds interpretations, this model is taken by assumption to give a maximally complete description. 
\item
(d) Models in which $\lambda$ includes sufficient information to reconstruct the wave function - this is more general than (c) above, and is what PBR aim to show must be the case, under additional assumptions.  This includes, e.g., the (deterministic) deBroglie-Bohm model, where the underlying properties required to predict all measurement outcomes are given by the position and the wave function (note, e.g., the outcome of a momentum measurement cannot be predicted from knowledge of the position alone). 
\end{list}

It is seen from the above that statistical completeness is satisfied by a wide class of models of interest, and hence is a relatively weak assumption.

Note that Eq.~(\ref{comp}) implies that information about the preparation - including information about the state that is prepared - may be lost.  In particular, one has a Markov chain, $P_\Psi \rightarrow \lambda \rightarrow (M=m)$, in which the last member is statistically independent of the first member.  The PBR result may be interpreted as showing that, under additional assumptions, no information about $\Psi$  itself is lost (although it is possible that all other information about the preparation procedure, e.g., the time it was switched on, cannot be reconstructed from $\lambda$). 

\subsection{Compatibility vs factorisability vs local compatibility}

\subsubsection{Compatibility}

It is natural, for any measurement $M$, to define a given property $\lambda$ as being {\it compatible} with preparation $P_{\Psi}$ if and only if there is a nonzero probability of the underlying properties being described by $\lambda$.  Writing $\lambda \sim (M,P_\Psi)$ to denote such compatibility, one thus has
\[ \lambda \sim (M,P_\Psi) ~~{\rm ~if ~and ~only ~if~~} p(\lambda|M, P_\Psi) > 0 . \]
Similarly, one may define $\lambda$ to be compatible with a given pure state $\Psi$ if and only if it is compatible with some preparation of $\Psi$.  

The `compatibility' assumption is simply the requirement that {\it if $\lambda$ is compatible with each of $\psi_1$, $\psi_2$,$\dots,\psi_n$, then it is compatible with $\psi_1\otimes\psi_2\otimes\dots\otimes\psi_n$}.   More formally: for any pure states $\psi_1$, $\psi_2$,$\dots,\psi_n$, there exist suitable preparation procedures $P_{\psi_1}$, $P_{\psi_2}$, ..., $P_{\psi_n}$ and $P_{\psi_1\otimes\psi_2\otimes\dots\otimes\psi_n}$ such that 
\begin{equation} \label{compat}
\lambda \sim (M,P_{\psi_j})~\forall j{\rm ~~implies~~} \lambda \sim (M,P_{\psi_1\otimes\psi_2\otimes\dots\otimes\psi_n}) .
\end{equation}

The compatibility assumption is natural, for example, in a first scenario where $n$ apparatuses, corresponding to the preparations $P_{\psi_1}$, $P_{\psi_2}$, ..., $P_{\psi_n}$, are each randomly chosen to either operate or not in a given instance, such that $\psi_1\otimes\psi_2\otimes\dots\otimes\psi_n$ is prepared when they happen to all be operating.  Thus, the probability $p(\lambda|M,P_{\psi_j})$ is conditioned only on the knowledge that the apparatus for procedure $P_{\psi_j}$ was operating, irrespective of the others. Hence, if $\lambda$ is compatible with each preparation procedure, irrespective of whether the other preparations procedures have been carried out, it is reasonable to expect it to be compatible with the case that all apparatuses are operating.

\subsubsection{Factorisability and reductionism}

PBR make a stronger `factorisability' assumption \cite{pbr}, which requires that 
\begin{list}{}{}
\item (A) the underlying properties $\lambda$ associated with a tensor product state $\psi_1\otimes\psi_2\otimes\dots\psi_n$ partition into respective properties $\lambda_1,\lambda_2,\dots,\lambda_n$ associated with each of $\psi_1$, $\psi_2$, $\psi_n$, i.e., $\lambda\equiv(\lambda_1,\lambda_2,\dots,\lambda_n)$; and 
\item (B) the underlying properties of the subsystems are uncorrelated.
\end{list}
Thus, 
\begin{equation} \label{fact}
    p(\lambda_1,\lambda_2,\dots,\lambda_n | M, P_{\psi_1\otimes\psi_2\otimes\dots\otimes\psi_n}) = \prod_{j=1}^n p(\lambda_j | M, P_{\psi_j}). 
\end{equation} 

One justification for factorisability is via a scenario in which apparatuses for preparing $\psi_1,\psi_2,\dots,\psi_n$ are operated simultaneously in spacelike separated regions, so that one may naturally associate local parameters, $\lambda_j$, with each preparation $P_{\psi_j}$. If it is then further assumed that these local parameters are uncorrelated, then Eq.~(\ref{fact}) will be satisfied.  A similar assumption has been recently used by Branciard {\it et al.} to derive a strengthened `bilocal' Bell inequality \cite{bilocal}.

Note, however, that in this second scenario the apparatuses share a common past, and hence it is consistent with causality for the $\lambda_j$ to be correlated, so that factorisability is violated.   In contrast, compatibility allows correlations between the $\lambda_j$, as long as whenever each $p(\lambda | M, P_{\psi_j})$ is positive, then $p(\lambda_1,\lambda_2,\dots,\lambda_n | M, P_{\psi_1\otimes\psi_2\otimes\dots\otimes\psi_n})$ is also positive. 

More generally, consider a scenario in which a partitioning of the underlying properties, as in (A) above, holds even if the respective preparations are simultaneously carried out in the same spacetime region.  This is the case for all `reductionist' models, where properties of systems are fully determined by properties of their subsystems.  Factorisability as per Eq.~(\ref{fact}) therefore requires that the apparatuses operate independently at the level of the underlying properties of the subsystems.
In contrast, in this third `reductionist' scenario, compatibility allows the possibility of correlations between the $\lambda_j$ - due, e.g., to direct physical interference, or to a common power supply.  

Thus, the compatibility condition can not only be formulated for nonreductionist models, but is far weaker than factorisability for the case of reductionist models.   Further, for {\it non}reductionist models, the compatibility assumption is reasonable to make in the `random operation' scenario described above.  Fortunately, this scenario corresponds to the experimental test proposed by PBR 
\cite{pbr}.

\subsubsection{Local compatibility for reductionist models}

For reductionist models, i.e., with $\lambda\equiv(\lambda_1,\lambda_2,\dots,\lambda_n)$, it is also of interest to consider an alternative assumption,  `local compatibility':
\begin{equation} \label{local}
\lambda_j \sim (M,P_{\psi_j})~\forall j{\rm ~~implies~~} \lambda \sim (M,P_{\psi_1\otimes\psi_2\otimes\dots\otimes\psi_n}) .
\end{equation}
This is clearly weaker than factorisability as per  Eq.~(\ref{fact}) -the only constraint imposed by local compatibility is that if each $\lambda_j$ is possible when the apparatus for $P_{\psi_j}$ is operated by itself, then $(\lambda_1,\lambda_2,\dots,\lambda_n)$ is  possible when the apparatuses are operated simultaneously.

Moreover, in contrast to the compatibility assumption (\ref{compat}), local compatibility only requires that the marginal probabilities, $p(\lambda_j|M,P_{\psi_j})$ are strictly positive on the left hand side of Eq.~(\ref{local}).  However, compatibility becomes equivalent to local compatibility if $p(\lambda|M,P_{\psi_j})$ only depends on $\lambda_j$.

For example, consider the case where apparatuses operate randomly, as in the first scenario above.  In this case it is natural to expect that $p(\lambda|M,P_{\psi_j})$, conditioned only on the operation of $P_{\psi_j}$, irrespective of the other apparatuses, is only dependent on $\lambda_j$, corresponding to the remaining components of $\lambda$ having a uniform distribution.  Integration over these remaining components then leads to $p(\lambda|M,P_{\psi_j})>0$ if and only if $p(\lambda_j|M,P_{\psi_j})>0$.  Thus, the compatibility assumption (\ref{compat}) becomes equivalent to local compatibility as per Eq.~(\ref{local}).

\subsection{Measurement independence - necessary or not?}

PBR further assume (e.g., throughout their Appendix B) that $\lambda$ is uncorrelated with the choice of measurement, i.e., that
\begin{equation} \label{meas}
    p(\lambda|M, P_\Psi) = p(\lambda| P_\Psi)
\end{equation}
(thus, e.g., $\mu_i(\lambda)$ in Eq.~(B1) of \cite{pbr} is independent of $M$).  This is the condition of `measurement independence', typically justified by an appeal to experimental free will \cite{relaxed}.  However, there is no reason, in principle, why one should not have a common-cause or direct-cause correlation between properties describing the world at some time, and the selection of a measurement setting at a later time - it is, after all, very natural to allow the past to influence the future.

It turns out that measurement independence is {\it not} required for obtaining a {\it limited} form of the PBR theorem, i.e., Theorem 1 in section 4 below.  This is essentially because the PBR approach only requires consideration of a single measurement procedure for each pair of pure states.  Further, it is only this limited form of the theorem that is relevant to the experimental test proposed by PBR, where only a single measurement procedure is required.

This reliance on a single measurement procedure is a unique feature of the PBR theorem.  The consequent lack of any  need to assume measurement independence, to obtain a limited form of the theorem, significantly differentiates it from other known `no-go' theorems for quantum phenomena, such as those based on Bell inequalities, Kochen-Specker theorems and steering inequalities.

However, an assumption of measurement independence, as per PBR, {\it is} needed to extend the limited form so as to obtain conclusions that are valid independently of any particular measurement (see Theorem 2 below).

\section{Generalising the PBR theorem}

\subsection{The main results and their significance}

A limited form of the PBR result can be obtained based only on equations (\ref{bayes})-(\ref{compat}) above, i.e, on Bayes theorem, statistical completeness, and compatibility:

{\bf Theorem 1:}  {\it The assumptions of statistical completeness and compatibility (or local compatibility) imply that, for a particular measurement $M$, there are scenarios for which the probability densities $p(\lambda | M, P_\psi)$ and $p(\lambda | M, P_\phi)$ are nonoverlapping for any two distinct pure states $\psi$ and $\phi$, with probability 1.}

Thus, given knowledge of the parameter $\lambda$, one can determine which one of the two nonoverlapping distributions $p(\lambda | M, P_\psi)$ and $p(\lambda | M, P_\phi)$ it is compatible with, and hence determine which of the two preparation procedures was carried out. Hence, the underlying properties of the system  allow scenarios in which one can, in principle, distinguish between $\psi$ and $\phi$ with certainty, even when these states are overlapping.  This is in marked contrast to the Helstrom bound for the minimum probability of error for distinguishing between $\psi$ and $\phi$, i.e., 
\[ P_e \geq \frac{1}{2} \left[ 1 - \sqrt{1-|\langle\psi|\phi\rangle|^2}\right]  . \]

Theorem 1 may be directly tested via the experimental scenario suggested by PBR \cite{pbr}  However, note that the simultaneous preparations in this experiment do {\it not} have to satisfy the factorisability assumption (\ref{fact}), but only the weaker compatibility assumption (\ref{compat}).  Thus, it is not necessary to assume a complete absence of correlations between the apparatuses - arising, e.g., from a common past, physical proximity or a shared power supply (see Section 3.2).  

It is remarkable that no assumption of measurement independence is needed to obtain Theorem 1 (and hence to interpret any experimental test thereof), in contrast to other `no go' theorems (see section 3.3).

The full form of the PBR theorem follows via the addition of measurement independence, yielding

{\bf Theorem 2:}  {\it The assumptions of statistical completeness, compatibility (or local compatibility) and measurement independence imply that there are scenarios for which the probability densities $p(\lambda | P_\psi)$ and $p(\lambda | P_\phi)$ are non-overlapping for any two distinct pure states $\psi$ and $\phi$,
with probability 1.}

In other words, for any pure quantum state, there are scenarios in which the wavefunction $\psi$ can be uniquely reconstructed from knowledge of $\lambda$ with probability unity.  This theorem corresponds to the original result proved by PBR \cite{pbr}, but with their factorisability assumption replaced by the more widely applicable and weaker assumption of compatibility.  It places a strong restriction on models of quantum phenomena, as discussed by PBR \cite{pbr} (see also Section 1 above).

Note the caveat about `probability 1' is necessary in the statement of the above theorems,  because $\lambda$ will typically be a continuous variable.  In particular, the theorems require $\lambda$ to contain information about the continuous variable $\psi$, where even for  qubits $\psi$ is parameterised by the surface of the Bloch sphere.  

Finally, note that to extend these theorems to be scenario-independent  would amount to requiring the distributions $p(\lambda|M,P_{\Psi})$ to be reproducible properties of the preparation procedure, independently of other physical phenomena such as the presence of other preparation apparatuses. This may be a reasonable assumption for particular classes of preparation procedures (those that are sufficiently stable for their operation to be independent of their environment), but will not be considered here.  

\subsection{The PBR measurement}

For any two distinct pure states $\psi$ and $\phi$, PBR determine a corresponding integer $n>1$, and construct a particular measurement $M$ which can be carried out on any tensor product of $n$ such states.  For the particular case where $|\langle\psi|\phi\rangle>|^2=1/2$ - e.g., the qubit states $|0\rangle$ and $ \frac{1}{\sqrt{2}}(|0\rangle+|1\rangle)$ - one has $n=2$, and $M$ is similar to a Bell state measurement, which is carried out on the four randomly prepared tensor product states $\psi\otimes\psi$, $\psi\otimes\phi$, $\phi\otimes\psi$, and $\phi\otimes\phi$ \cite{pbr}.

More generally there are precisely $2^n$ such tensor product states which can be formed from copies of $\psi$ and $\phi$, ranging from $\Psi_1=\psi\otimes\psi\otimes\dots \otimes\psi$ (i.e., $n$ copies of $\psi$) to $\Psi_{2^n}=\phi\otimes \phi\otimes \dots\otimes \phi$ (i.e., $n$ copies of $\phi$).  Each tensor product comprises $r$ copies of $\psi$ and $n-r$ copies of $\phi$, in some order, for some $r\in\{0,1,\dots n\}$.  

Further, the measurement $M$ is cleverly constructed  by PBR to have $2^n$ possible outcomes, $m=1,2,3,\dots,2^n$, with the following remarkable property:
\begin{equation} \label{prop} 
p(m|M,P_{\Psi_k}) = \langle \Psi_k|M_m|\Psi_k\rangle = 0 {\rm~~for~~}m=k  .
\end{equation}
The construction of such a measurement is a highly nontrivial exercise \cite{pbr},  and is the key to the PBR theorem and to Theorems 1 and 2 above.

\subsection{Proof of Theorem 1}

First, note that substitution of measurement property (\ref{prop}) into Eq.~(\ref{bayes}), for the case $k=m$, gives
$\int d\lambda\, p(m|M,P_{\Psi_m},\lambda)\,p(\lambda|M,P_{\Psi_m}) = 0$
Hence, since the quantity being integrated is nonnegative, integration over any subset $S$ of parameters must also vanish, i.e.,
\begin{equation} \label{zero}
\int_{S} d\lambda\, p(m|M,P_{\Psi_m},\lambda)\,p(\lambda|M,P_{\Psi_m}) =0 
\end{equation}

Choosing $S$ to be the subset defined by
\begin{equation} \label{sdef}
S := \{\lambda: \lambda\sim (M,P_\psi),~\lambda\sim (M,P_\phi)\} = \{\lambda:p(\lambda|M,P_\psi)\,p(\lambda|M,P_\phi)>0 \}, 
\end{equation}  
the compatibility assumption (\ref{compat}) immediately implies that the preparations can be chosen such that the second term in the integral of Eq.~(\ref{zero}) is strictly positive. Hence,  $p(m|M,P_{\Psi_m},\lambda)=0$
for all $\lambda \in S$, except possibly on some zero measure subset $S_m$ of $S$. 
From the `completeness' assumption (\ref{comp}) it then follows that
\begin{equation} \label{first}
p(m|M,\lambda)=0 
\end{equation}
for all $\lambda\in S$, except possibly on some zero measure subset $S_m$.  

Eq.~(\ref{first}) thus holds for all $2^n$ measurement outcomes $m$, except possibly on a zero measure subset $\cup_mS_m$ of $S$. But, for any $\lambda \in S\backslash \cup_mS_m$, summing this equation over  $m$  gives the contradiction $1=0$.  
Hence, $S=\cup_mS_m$, i.e, $S$ has zero measure. If follows immediately via Eq.~(\ref{sdef}) that
\begin{equation} \label{over}
\int d\lambda\, p(\lambda|M,P_\psi)\,p(\lambda|M,P_\phi) = \int_S d\lambda\, p(\lambda|M,P_\psi)\,p(\lambda|M,P_\phi) =0. 
\end{equation}
Thus, $p(\lambda|M,P_\psi)\,p(\lambda|M,P_\phi)=0$ except possibly on a set of measure zero, as per Theorem 1.

If compatibility is replaced by local compatibility, then $S$ is replaced in the above by the set of $(\lambda_1,\lambda_2,\dots,\lambda_n)$ for which $\lambda_1=\lambda_2=\dots=\lambda_n=\lambda_{\rm local}$ and $p(\lambda_{\rm local}|M,\phi)\,p(\lambda_{\rm local}|M,P_\phi)>0$.  This case closely corresponds to the proof of the PBR theorem in \cite{pbr}, but does not rely on factorisability.

\subsection{Proof of Theorem 2}

This is a straightforward corollary of Theorem 1, as the measurement independence assumption (\ref{meas}) implies that probability distributions for $\lambda$ are independent of the particular measurement $M$, thus extending Eq.~(\ref{over}) to all measurements.

{\bf ACKNOWLEDGEMENTS}

I am grateful to T. Rudolph, M. Pusey and H. Wiseman for several helpful comments. This research was supported by the ARC Centre of Excellence CE110001027.

\end{document}